\documentclass[fleqn,usenatbib]{mnras}

\usepackage{newtxtext,newtxmath}
\usepackage[T1]{fontenc}

\DeclareRobustCommand{\VAN}[3]{#2}
\let\VANthebibliography\thebibliography
\def\thebibliography{\DeclareRobustCommand{\VAN}[3]{##3}\VANthebibliography}

\usepackage{graphicx}	% Including figure files
\usepackage{amsmath}	% Advanced maths commands
\usepackage{breakurl}
\usepackage{color}
\usepackage{pdflscape}
\usepackage{multirow}
\usepackage{mathtools}  % for multilines in eq.1
\usepackage{longtable}
\usepackage{siunitx}    % for alignment by \pm in table A2
\newcommand{\dg}{^{\circ}}
\newcommand{\rtwofact}{1.58}

\newcommand{\rthreefact}{0.56}

\newcommand{\rfourfact}{2.51}

\title[Repeated pattern of gamma-ray flares in 3C~279]{Repeated pattern of gamma-ray flares in the lightcurve of the blazar 3C~279}

\author[D. Blinov et al.]
{D. Blinov$^{1,2,3}$\thanks{E-mail: blinov@ia.forth.gr},
S.~G.~Jorstad$^{4,3}$,
V.~M.~Larionov$^{3}$\thanks{Deceased}, N.~R.~MacDonald$^{5}$, T.~Grishina$^{3}$,\newauthor E.~Kopatskaya$^{3}$, 
E.~Larionova$^{3}$, L.~Larionova$^{3}$, D.~Morozova$^{3}$, A.~Nikiforova$^{3,6}$, 
\newauthor
S.~Savchenko$^{3,6,7}$, Y.~Troitskaya$^{3}$, I.~Troitsky$^{3}$\\
$^{1}$Institute of Astrophysics, Foundation for Research and Technology-Hellas, Voutes, 71110 Heraklion, Greece\\
$^{2}$Department of Physics, 
University of Crete, 71003, Heraklion, Greece\\
$^{3}$Astronomical Institute, St. Petersburg State University, Universitetsky pr. 28, Petrodvoretz, 
198504 St. Petersburg, Russia \\
$^{4}$Institute for Astrophysical Research, Boston University, 725 Commonwealth Avenue, Boston, MA02215\\
$^{5}$Max-Planck-Institut f{\"u}r Radioastronomie, Auf dem H{\"u}gel 69, 53121 Bonn, Germany\\
$^{6}$Pulkovo Observatory, 196140, St.Petersburg, Russia\\
$^{7}$Special Astrophysical Observatory, Russian Academy of Sciences, 369167, Nizhnii Arkhyz, Russia\\
}

\date{Accepted 2021 May 20. Received 2021 May 20; in original form 2021 March 15}

\pubyear{2021}

\begin{document}
\label{firstpage}
\pagerange{\pageref{firstpage}--\pageref{lastpage}}
\maketitle

\begin{abstract}
The optical polarization plane of some blazars occasionally exhibits smooth hundred degree long rotations. Multiple theoretical models have been proposed to explain the nature of such events. A deterministic origin of these rotations, however, remains uncertain. We aim to find repeating patterns of flares in gamma-ray light curves of blazars, which accompany optical polarization plane rotations. Such patterns have been predicted to occur by one of the models explaining this phenomenon. For the blazar 3C~279, where multiple polarization plane rotations have been reported in the literature, we obtain the {\em Fermi}-LAT gamma-ray light curve and analyze its  intervals adjacent to polarization plane rotations. We find a complex characteristic pattern of flares in the gamma-ray light curve that is repeated during periods adjacent to three large amplitude EVPA rotation events in 3C~279. We discover a "hidden EVPA rotation", which can only be seen in the relative Stokes parameters plane and that occurred simultaneously with the fourth repetition of the pattern. This finding strongly favors the hypothesis of emission features propagating in the jet as the reason of optical polarization plane rotations. Furthermore, it is compatible with the hypothesis of a sheath in the jet comprised of more slowly propagating emission features.
\end{abstract}

\begin{keywords}
gamma-rays: galaxies -- galaxies: jets -- galaxies: nuclei -- polarization -- quasars: individual: 3C 279
\end{keywords}

%%%%%%%%%%%%%%%%%%%%%%%%%%%%%%%%%%%%%%%%%%%%%%%%%%

%%%%%%%%%%%%%%%%% BODY OF PAPER %%%%%%%%%%%%%%%%%%

\section{Introduction} \label{sec:introduction}

Blazars are active galactic nuclei (AGN) with relativistic jets pointed nearly towards the Earth. This alignment causes strong relativistic boosting of the jet emission \citep{Blandford1979}, which leads to peculiar observational properties of blazars. For instance, the spectral energy distribution (SED) of blazars has two broad peaks. They are typically located in the IR to X-ray frequencies and in the gamma-ray band. It is well established that the emission of the low energy peak is produced by relativistic electrons in the jet emitting synchrotron radiation. The high energy emission of blazars is commonly attributed to inverse Compton emission. However, physical processes underlying the gamma-ray emission of blazars remain an active area of research in astrophysics.

The synchrotron origin often causes high polarization in the optical emission of blazars, and typically exhibits erratic variations of the electric vector position angle (EVPA) and fractional polarization \citep[e.g., ][]{Uemura2010}. Nevertheless, a number of events have been observed in different blazars, where the polarization plane in the optical exhibited gradual hundreds degrees long rotations. It has been argued that these EVPA rotations are related
to high energy flares \citep[e.g.,][]{Marscher2008,Abdo2010,Aleksic2014b}. However, the connection between gamma-ray flares and EVPA rotations has not been firmly established. Moreover, the physical mechanisms behind these events remain unclear. There are a handful of theoretical models proposed to explain these EVPA rotations \citep[see][and the references therein]{Blinov2019}. Most of these models successfully reproduce the general properties of EVPA rotations such as duration and amplitude \citep[see e.g.][]{Kiehlmann2017,Peirson2019,Hosking2020,Cohen2020}. Therefore, more sophisticated observables are required in order to discriminate among these models.

Here we report the first discovery of a persistent pattern of gamma-ray flares accompanying EVPA rotations in the emission of the blazar 3C~279. Possible existence of such patterns has been predicted by \cite{Marscher2010}, who reported on a series of gamma-ray flares in the blazar PKS~1510$-$089 during its optical EVPA rotation culminating in a new radio knot ejection at 43~GHz. Their interpretation of the events included an emission feature propagating along a curved trajectory in the acceleration and collimation zone of the jet with a helical magnetic field. This emission feature produced a series of synchrotron self-Compton (SSC) and external-Compton (EC) gamma-ray flares during its propagation. The EC flares have been explained by an external photon field produced by other emission features in the slower sheath of the jet. \cite{Marscher2010} concluded: "The relatively slow motion and gradual evolution of such features implies that they should persist for years, in which case another series of flares in the near future should exhibit a similar pattern of variability and appearance of a superluminal knot.". In this paper we demonstrate that such persistent patterns of gamma-ray flares are indeed exhibited by blazars and they are inherently related to concurrent EVPA rotations.

The values of the cosmological parameters adopted throughout this work are $H_0 = 67.7$ km s$^{-1}$ Mpc$^{-1}$, $\Omega_m = 0.307$, and $\Omega_\Lambda = 1 - \Omega_m$ \citep{Planck2016}.

\section{Observations and data reduction} \label{sec:obs}

\subsection{Optical data} \label{subsec:opt_data}

The optical data were collected from multiple telescopes and observing programs. In the Julian date (JD) range between 2454806 and 2456133 (2008 December 5 -- 2012 July 24) we used optical polarimetry and R-band photometry data from \cite{Kiehlmann2016} excluding points marked as outliers there, moreover, we excluded measurements with uncertainties of R magnitude $> 0.09^m$. In the ranges $2454560 \le JD \le  2454806$ (2008 April 3 -- 2008 December 5) and $2456133 \le JD \le 2457820$ (2012 July 24 -- 2017 March 7) we used photometric and polarimetric measurements from the Crimean observatory 70 cm AZT-8, the St. Petersburg  University  40~cm LX-200 telescopes \citep{Larionov2008}, the
Perkins 1.82 m telescope (PTO, Flagstaff, AZ), and publicly available data of the 1.54~m Kuiper and the 2.3~m Bok telescopes of the Steward observatory 
\citep{Smith2009}. The photometric measurements have been corrected for the Galactic extinction following \cite{Schlafly2011}. We accounted for the 180$^\circ$ ambiguity of EVPA minimizing the consecutive changes of the curve, i.e. the following value $|EVPA_{\rm i} - EVPA_{\rm i-1}|$ \citep[see e.g.][]{Kiehlmann2016}.

\subsection{Gamma-ray data} \label{subsec:gamma_data}

For the gamma-ray data we analysed the {\em Fermi} Large Area Telescope (LAT) data. The {\em Fermi} gamma-ray space observatory observes the entire sky at energies of 20 MeV -- 300 GeV every three hours \citep{Atwood2009}. We processed the data in the energy range $100\, {\rm MeV} \le E \le 300\, {\rm GeV}$ using the unbinned likelihood analysis of the standard {\em Fermi} analysis software package Fermitools (v. 1.2.23) distributed under Conda. Throughout the paper we used light curves with 1.35 days binning. However, we also verified that all results hold for two other values of binning 1.4 and 2 days. We used the instrument response function $P8R3\_SOURCE\_V2$. Source class photons (evclass=128 and evtype=3) were selected within a $15\dg$ region of interest (ROI) centered on 3C 279. A cut on the satellite zenith angle ($< 90\dg$) was enabled to exclude the Earth limb background. The Galactic interstellar emission was accounted for using the $gll\_iem\_v07$ spatial model. The extragalactic diffuse and residual instrumental backgrounds were included in the fit as an isotropic spectral template $iso\_P8R3\_SOURCE\_V2\_v1.txt$. The background models include all sources from the 4FGL \citep[fourth {\em Fermi}-LAT source catalogue,][]{Abdollahi2020} within $15\dg$ of the blazar. Photon fluxes of sources beyond $10\dg$ from 3C 279 and spectral shapes of all targets were fixed to their values reported in 4FGL. The source is considered to be detected if the test statistic, TS, provided by the analysis exceeds 10, which corresponds to approximately a $3\sigma$ detection level \citep{Nolan2012}. Since the Sun passes in a close vicinity of 3C 279, we verified that its emission does not affect derived photon fluxes significantly. We processed two intervals of the light curve in 2009 and 2010, when the Sun disk was within the $15\dg$ ROI accounting for its stationary emission with a template calculated using the Solar System Tools \citep{Johannesson2013}. Deviation of these intervals of the light curve from identical intervals without accounting for the stationary Sun emission was found to be smaller than the point size in plots of this paper, i.e. negligible.

\section{Flares pattern during EVPA rotations} \label{sec:res}
\begin{figure*}
   \centering
   \includegraphics[width=0.97\textwidth]{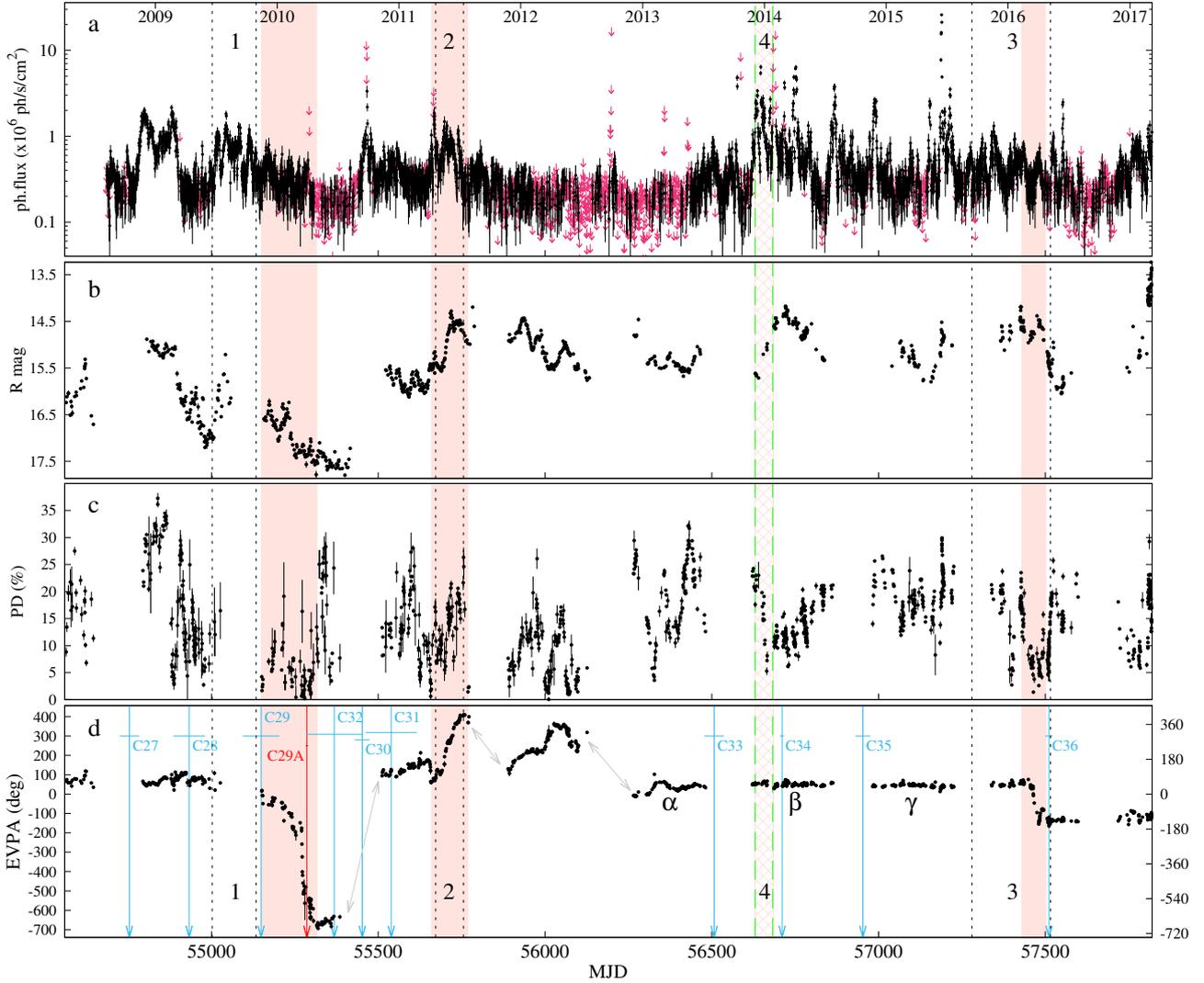}
      \caption{Panel \textbf{a}: {\em Fermi LAT} gamma-ray photon flux curve in logarithmic scale with 1.35 days binning. Panel \textbf{b}: R-band optical light curve. Panel \textbf{c}: optical fractional polarization. Panel \textbf{d}: optical polarization position angle. The filled pink areas mark periods of three largest EVPA rotations. The hatched pink area indicates period of a hidden EVPA rotation. The vertical cyan arrows show the moments of new radio knot ejections at 43 GHz from \protect\cite{Larionov2020}. The red arrow indicates the moment of C29 knot ejection if only its first three epochs are considered (see Sec.~\ref{sec:analysis}). The gray arrows indicate seasonal gaps, where the EVPA curve was shifted by $\pm n \times 180\dg$ $(n \in \mathbb{N})$ in order to keep each season beginning in a narrow range of values. The numbered intervals between the vertical dashed lines are the regions when the gamma-ray curve shows the repeated pattern. Zoomed versions of these intervals are shown in Figs.~\ref{fig:g_patt}, \ref{fig:pat4} and \ref{fig:origgamma}.}
         \label{fig:all}
\end{figure*}

The collected data set is presented in Fig.~\ref{fig:all}, where the pink regions denote time intervals of the three largest EVPA rotations observed during the reported period.
These rotation events have been reported and analyzed in previous works \citep{Kiehlmann2016,Aleksic2014,Larionov2020}.

\begin{figure*}
   \centering
   \includegraphics[width=0.90\textwidth]{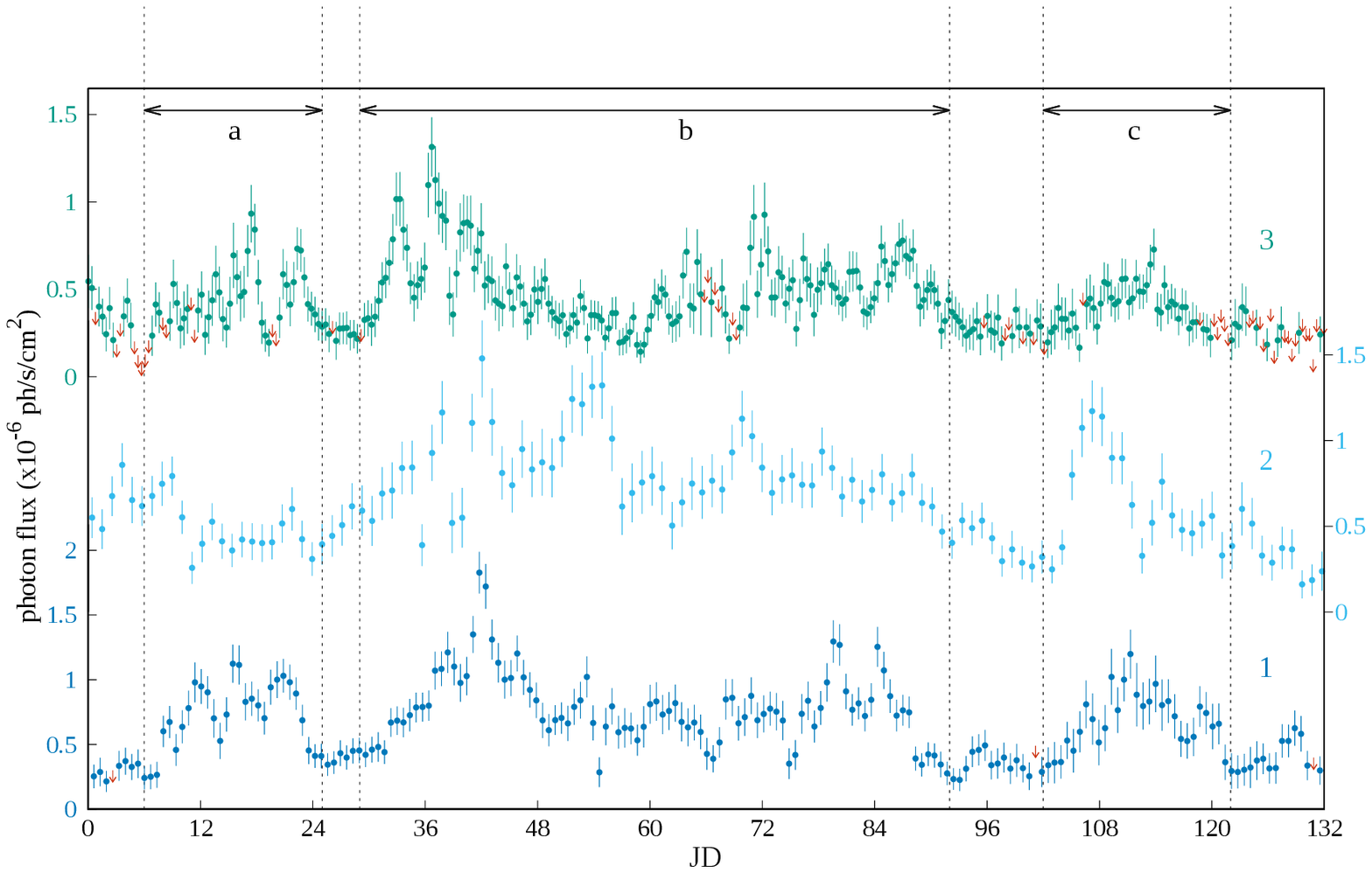}
      \caption{Fermi LAT gamma-ray light curves during periods denoted with black dashed lines in Fig.~\ref{fig:all}. JD=0 corresponds to MJD=55002 for the first curve. The curves are color-coded to denote three successive flaring events. The integration time for each point is 1.35d. The curves are over-sampled in such a way that the distance between consecutive points equals to half of the integration time. The vertical dashed lines indicate three main sets of flares. Curve 1 has its original time scale. The time axes for two other curves is adjusted by factors $\rtwofact$ and $\rthreefact$ for curves 2 and 3 respectively. A similar plot with unaltered time axes is given in Appendix~\ref{app:A}.}
         \label{fig:g_patt}
\end{figure*}

Visually comparing parts of the gamma-ray light curve along the EVPA rotations we have found that they closely resemble each other during the time intervals 1 -- 3 shown by the vertical dashed lines in Fig.~\ref{fig:all}. Since the jet Doppler factor may vary between different events affecting the time dilation, we found the best agreement between the corresponding pieces of the light curve by stretching or squeezing them. To this end we minimized the Time-Warp Edit Distance \citep[TWED][]{Lin2012} between intervals of the light curve as a function of a multiplication factor of the time axis for one of the intervals. TWED is an elastic distance metric that can account
for irregular or inconsistent time intervals between points in compared light curves. We found that the best agreement between the curve parts is achieved when the time scale in intervals 2 and 3 is multiplied by a factor of $\rtwofact$ and $\rthreefact$ respectively. In Fig.~\ref{fig:g_patt} we show the three intervals of the light curve after the stretching/compression and shifting to the time range of the first interval. It is remarkable that all three parts represent very similar complex pattern of flares. In Appendix \ref{app:B} we present evidence that interval 1 indeed provides the best match along the entire gamma-ray light curve with intervals 2 and 3 after the corresponding transformations of their time scales. Despite of differences in the fine light curve structure one can distinguish 3 major flares sequences that are outlined by the dashed lines in Fig.~\ref{fig:g_patt} and denoted $a$, $b$ and $c$. Sequence $a$ is missing in the curve 2, however, the two other sequences $b$ and $c$ resemble corresponding events in curves 1 and 3 in the amplitude, duration and relative position on the adjusted time axis.

\section{Hidden EVPA rotation} \label{sec:hid}

\begin{figure*}
   \centering
   \includegraphics[width=0.7\textwidth]{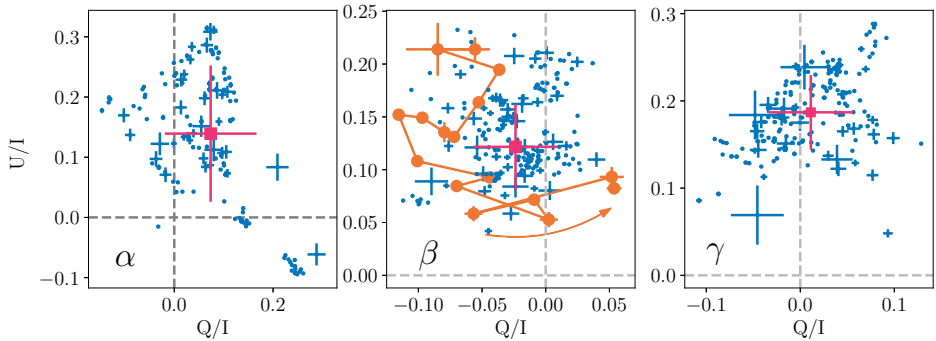}
      \caption{Distributions of measurements on the relative Stokes parameters Q/I - U/I plane for the three seasons marked $\alpha$, $\beta$ and $\gamma$ in Fig.~\ref{fig:all} are shown by blue and orange points. The orange points represent daily averaged consecutive measurements in the range 56632 < MJD < 56687, which is marked by the hatched pink region in Fig.~\ref{fig:all}. The magenta squares indicate centroid of each distribution calculated as a sigma-clipped median, while its uncertainties are the standard deviations of all measurements along corresponding axes.}
         \label{fig:qu}
\end{figure*}

During the time interval between 2012 October and 2015 September ($56220 < {\rm MJD} < 57290$) the EVPA curve in Fig.~\ref{fig:all} shows almost no variability compared to three previous observing seasons. At the same time the fractional polarization during this period is as variable as it was before, but has higher average value. The higher level of polarization explains the lack of variability of EVPA, which becomes evident when the polarization measurements are plotted on the relative Stokes parameters plane. Such plots for the three seasons marked as $\alpha$, $\beta$ and $\gamma$ in Fig.~\ref{fig:all} are presented in Fig.~\ref{fig:qu}.
For each of the three observing seasons centroid of the distribution on the Q/I - U/I plane is significantly shifted from the origin of coordinates. Therefore, values of EVPA\footnote{that is by definition 1/2 of the angle between the positive direction of the Q/I axis and the vector connecting (0,0) with the measurement position on the Q/I - U/I plane.} vary in a small range $\pm 30 - 35\dg$ around the value corresponding to the centroid of each distribution. For this reason loops of consecutive measurements on the Q/I - U/I plane, if present, cannot be seen as rotations in the EVPA curve in principle, because they do not encompass (0,0). Such behaviour of the polarization vector can be called "hidden EVPA rotations" and it is common for blazars \cite[e.g.][]{Ikejiri2011,Larionov2016}. It is caused by a quasi-stationary strongly polarized component present in the optical emission \citep{Villforth2010,Uemura2010}.

In order to identify loops on the relative Stokes parameters plane during three observing seasons $\alpha$ - $\gamma$ we used TimeTubesX\footnote{\url{https://timetubes.herokuapp.com/}} web-application that was created specifically for this purpose \citep{Uemura2017,Fujishiro2018}. With its help we found a sequence of consecutive measurements between 2013 December 6 and 2014 January 29 that constitute 3/4 of a loop on the relative Stokes parameters plane shown in the middle panel of Fig.~\ref{fig:qu}. This loop would be seen as a $\sim120\dg$ rotation in the EVPA curve, if the origin of coordinates was shifted to the centroid of the distribution shown by the magenta point. The latter operation is equivalent to subtraction from the total emission  of the highly polarized underlying component that hides the EVPA rotation. The time interval when this hidden EVPA rotation occurred is marked by the pink hatched area in Fig.~\ref{fig:all}.

\begin{figure*}
   \centering
   \includegraphics[width=0.9\textwidth]{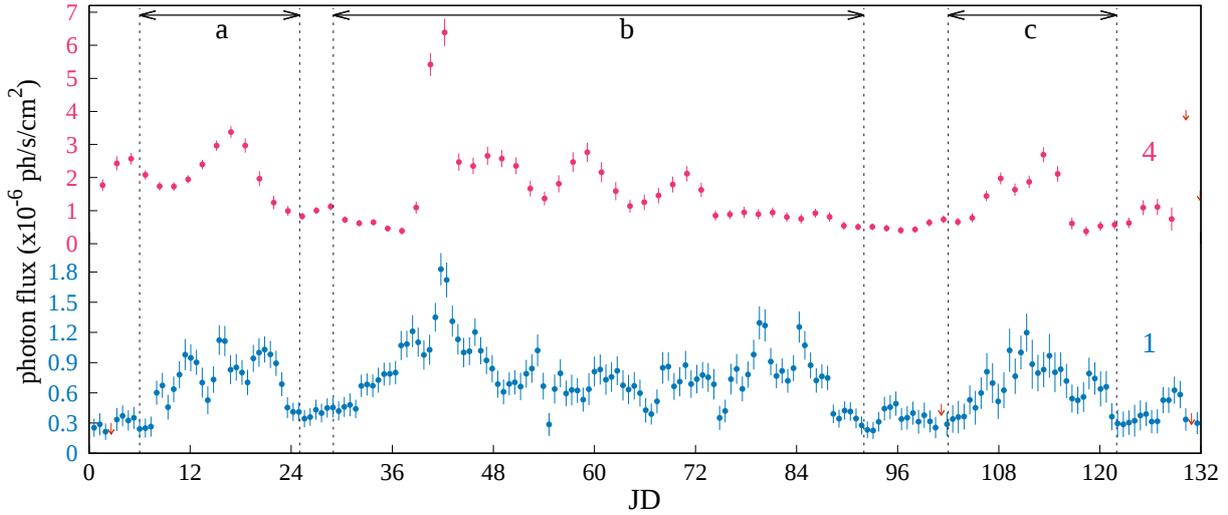}
      \caption{Fermi LAT gamma-ray light curve during the period marked with green dashed lines in Fig.~\ref{fig:all}, which is nearly coincident with the hidden EVPA rotation is shown in magenta. It is stretched by a factor of $\rfourfact$ and demonstrated together with the first episode of the repeated pattern (blue) that is shown in its original time scale. As in Fig.~\ref{fig:g_patt} JD=0 corresponds to MJD=55002 for the first curve. The integration time for each point is 1.35d. The curves are over-sampled in such a way that the distance between consecutive points equals to half of the integration time. The black dashed lines indicate three main sets of flares.}
         \label{fig:pat4}
\end{figure*}

We compared the gamma-ray light curve during the time interval adjacent to the hidden rotation with the repeated pattern of flares found for the three evident rotations. Using the same minimization procedure as described in the previous section we found that the gamma-ray light curve in the range $56630.1 < MJD < 56682.7$ (marked by green dashed lines in Fig.~\ref{fig:all}) after scaling it by a factor of $\rfourfact$ provides the best match to interval 1 of the repeated pattern. In Fig.~\ref{fig:pat4} we show this interval of the gamma-ray light curve stretched $\rfourfact$ times along the time axis together with interval 1 in its original scale for comparison. One can note that despite the differences in amplitudes both intervals resemble each other in relative flare amplitude and their duration. We speculate that this part of the gamma-ray light curve represents the fourth repetition of the gamma-ray flares pattern found in the previous section. This time it occurred essentially simultaneously with a hidden anticlockwise EVPA rotation that can only be seen on the  Stokes parameters plane.

\section{Flares mechanism and emission zone} \label{sec:analysis}

The discovered pattern of gamma-ray flares has been repeated 3 times during the time intervals adjacent or coincident with the largest amplitude rotations in the EVPA curve. With a Monte-Carlo simulation we estimated the probability of chance coincidence of this sequence of events as $< 5.3 \times 10^{-5}$ ($> 4\sigma$ significance). If the fourth repetition of the pattern during the hidden EVPA rotation is also considered in this simulation then the chance coincidence is reduced to $< 3.1 \times 10^{-6}$ ($> 4.6\sigma$ significance). This fact by itself essentially rules out the possibility that the pattern is just a random process that is unrelated to the optical polarization variability.

\begin{figure}
   \centering
   \includegraphics[width=\hsize]{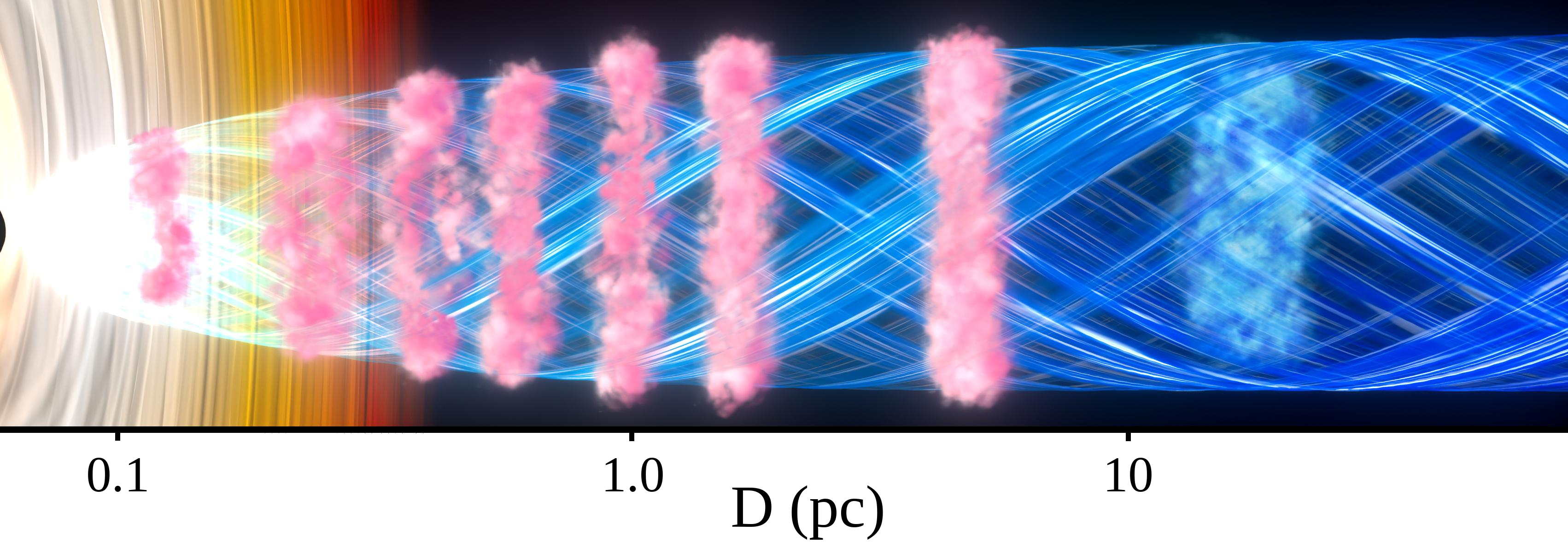}
      \caption{Cartoon of the model for 3C 279 discussed in the text. The logarithmic length scale is used for demonstration of different AGN components. The transversal scale is different from the longitudinal. Blue plasma corresponds to the fast spine of the jet and the emission region propagating within it. Pink regions are the ring-like condensations in the sheath that provide the external photon field for the External Compton process taking place in the moving emission feature.}
         \label{fig:jetprof}
\end{figure}

The most plausible interpretation of the observed set of events is the model proposed by \cite{Marscher2010}. Each of the three episodes can be explained by an emission feature propagating along a curved trajectory in the fast jet spine permeated with a helical magnetic field. We posit that as the emission feature propagates through the acceleration and collimation zone of the jet, it passes through several quasi-stationary emission regions associated with seed photons emanating from a slower outer jet sheath \citep{Ghisellini2005,MacDonald2017}. This model is depicted in Fig.~\ref{fig:jetprof}.
Optical and/or infrared seed photons from these regions are up-scattered to gamma-ray energies due to the external Compton (EC) mechanism \citep{MacDonald2015}, which causes the set of gamma-ray flaring events $a$ - $c$. The EC nature of the gamma-ray flares in 3C~279 for the considered periods has been demonstrated by \cite{Larionov2020}. The moving emission feature presumably fills only a part of the spine cross-section, therefore, it traces the local magnetic field direction that changes along the trajectory. Since its relativistically boosted radiation dominates in the optical band at least during some part of the emission feature lifetime, we observe the optical EVPA rotation during its propagation down the jet.

Within the framework of the proposed interpretation we can estimate the distance traveled by the emission features during the set of flares $a$ -- $c$ using parameters of radio knots observed at 43 GHz with VLBI (shown by arrows in Fig.~\ref{fig:all}). Association of individual radio knots with the emission features responsible for events 1 - 4 is not straightforward. Regions 3 and 4 can be associated with knots C36 and C34 fairly reliably, since these knots passed the radio-core right after the end of the corresponding pattern and EVPA rotation. Similar behaviour was observed by \cite{Marscher2010} in PKS1510-089, where the set of flares occurred before the radio-knot ejection. This is consistent with upstream location of the emission zone with respect to the radio-core. Among knots reported by \cite{Jorstad2017} and \cite{Larionov2020} there are no possible associations for region 2 that were ejected soon after it. However, one may notice in Fig.~\ref{fig:all} that there is a large gap between ejection time of knots C31 and C33. In the range $55539 < MJD < 56507$ there were no new knots identified in 43 GHz radiomaps. If ejection of knots was a Poisson process with the average frequency of events corresponding to that in the range $54754 \le MJD \le 55539$, when 6 knots C27 - C32 were ejected, then the probability of observing zero new knots in the following 968 days would be $\sim 6\times10^{-4}$. Therefore, it is unlikely that the absence of radio-knots between C31 and C33 is accidental. We attribute this lack of new knots ejections to the exceptional brightness of knots C31 and C32 that were propagating along this time interval within 0.7 mas from the core. During this period C32 was comparable with the core in total flux density, while C31 was several times more luminous \cite[see Fig. 14 in][]{Larionov2020}. Therefore, we cannot exclude the possibility that a knot associated to the repeated pattern region 2 was overlooked. For region 1 there are two possibilities of association. Due to the large uncertainty of ejection time C32 is consistent with the end of the EVPA rotation. On the other hand, trajectories of knots in 3C~279 are strongly non-ballistic \citep{Jorstad2017}. Therefore, derived ejection times and other parameters of knots depend on the considered period. If only first three epochs for knot C29 are fitted instead of 8 analyzed in \cite{Larionov2020}, then its ejection time is shifted to ${\rm MJD}=55286\pm4$. This value coincides with the end of the EVPA rotation of period 1 as demonstrated by the red arrow C29A in Fig.~\ref{fig:all}. We, therefore, cannot rule out the possibility that event 1 is in fact associated with C29 rather than C32. In any case the knot ejection is likely to be associated with the end of the EVPA rotation as in two other events.

Despite missing certain individual associations, we assume that the four emission features producing the patterns of flares and EVPA rotations have on median same velocities as the radio knots. During the observing period discussed in this paper there are 10 radio-knots C27 -- C36 (see Fig.~\ref{fig:all}) ejected from the core with median proper motion $<\mu> = 0.40\pm0.03$ mas yr$^{-1}$ and viewing angles from $\le 2\dg$ to $6\dg$ as reported by \cite{Larionov2020}. Therefore, adopting the average viewing angle $\sim 4 \dg$ we find that a typical emission feature must have traveled $11$ pc in the jet during the 108 days median span of the gamma-ray flares pattern. Even adopting proper motion of the slowest radio-knot $\mu_{\rm min} = 0.168\pm0.016$ mas yr$^{-1}$ (C32 in \cite{Larionov2020}) for the shortest pattern region 4 lasting 53 days and maximum viewing angle 6$\dg$ we find the lower limit on the distance traveled in the jet $3.7\pm0.3$ pc. Following this estimate, the only viable scenario appears to be the jet sheath as the source of seed photons for the EC mechanism producing the observed gamma-ray variability $a$ -- $c$. Because other possible sources of seed photons: the broad line region and the dust torus have substantially smaller size than the estimated distance traveled by the emission feature between flares $a$ and $c$.

The time scaling factors $\rtwofact$, $\rthreefact$ and $\rfourfact$ that provide the best match of regions 2, 3, 4 of the gamma-ray light curve with region 1 in the proposed scenario correspond to ratios of the Doppler factors $\delta_2/\delta_1$, $\delta_3/\delta_1$ and $\delta_4/\delta_1$ of the moving emission features. Since the observed flux density of optical synchrotron emission and EC gamma-ray emission depend on the Doppler factor as $F_{\rm syn} \propto \delta^{3+\alpha}$ and $F_{\rm EC} \propto \delta^{4+2\alpha}$ \citep{Ghisellini2005}, the ratio $F_{\rm EC}/F_{\rm syn}$ is $\propto \delta^{1+\alpha}$, where $\alpha$ is the spectral index of the emission \citep{Dermer1995}. We found that ratios of gamma-ray and optical fluxes in regions 1 and 2 near the peak of the first flare in event $b$ imply the $\delta_2/\delta_1 = 1.2\pm0.1$. For the decaying part of the same flare, where we have optical data in the region 3 we find $\delta_3/\delta_1 = 0.7\pm0.1$. Similarly, for the peak of the corresponding flare in region 4, where we linearly interpolated nearby optical measurements $\delta_4/\delta_1 = 2.2\pm0.1$. Given the fast variability, very different integration time in the two bands and possible presence of multiple emitting components these values provide a rather good agreement with ratios of Doppler factors derived from changes of the timescale of events in the gamma-ray curve.

\section{Discussion} \label{sec:disc}

A repeated pattern of gamma-ray flares was previously reported by \cite{Jorstad2013} in another blazar 3C 454.3, where it occurred three times during 2009 -- 2010. There the authors interpret these events by a presence of a system of standing conical shocks or magnetic reconnection events located in the parsec-scale millimeter-wave core of the jet. Similar shock-shock interaction between a traveling and a standing shock waves was proposed for other sources \citep[e.g., ][]{Fromm2013}. However, such interpretation is rather implausible for the events presented in this paper due to the EC nature of gamma-ray flares in 3C~279 demonstrated by \cite{Larionov2020}, which is inconsistent with the interacting shocks scenario. Moreover, \cite{Shukla2020} have recently discussed shorter duration (minutes timescale) events and much simpler patterns of peak-in-peak flares in 3C 279. These events could be characteristic of magnetic reconnection events. However, they have several orders of magnitude shorter durations and correspond to much smaller spatial scales in the jet compared to the events in this work. We emphasize that we do not attribute {\em all} gamma-ray activity in 3C~279 to the EC scattering of seed protons from the sheath emission regions discussed along the paper. It is likely that substantial fraction of flares in the gamma-ray light curve is produced by other mechanisms including the shock-shock interactions and the magnetic reconnections mentioned above.

It is worth noting that the same scenario adopted here for interpretation of the repeated pattern was used for explanation of "orphan" gamma-ray flares in blazars \citep{MacDonald2015,MacDonald2017}. In fact, the brightest gamma-ray flare of region 2 occurred during a rather low and inactive state in the optical and can be considered as an "orphan" flare. In three other regions due to sparse optical observations it is hard to say if any flares in the pattern miss optical counterparts. However, the possibility of such association can be used for identification of repeated patterns in other blazars. It has been found that $\sim 20$ percent of gamma-ray flares in blazars are orphan \citep{Liodakis2019}. Some of them can be potentially related with repeated patterns of flares similar to the one discussed in this work.

If persistent patterns of gamma-ray flares similar to the one found in this work are commonly present in the light curves of other blazars, they can be of interest for testing hypotheses related to propagation of plasma in the acceleration and collimation zone.
Here we discuss only one immediate implication of the finding concerning optical EVPA rotations in blazars. The fact that the complex pattern of gamma-ray flares repeated three times during or next to time intervals when the highest amplitude EVPA rotations occurred resolves two questions commonly discussed in blazar studies. First, the combination of the events unambiguously indicates that some EVPA rotations in blazars are deterministic events. Second, these EVPA rotations are indeed connected to gamma-ray flares. Currently, the model proposed by \cite{Marscher2010} best explains the sequence of events reported here. However, we do not claim that {\em all} EVPA rotations in blazars are explained by the same process as discussed here. There are EVPA rotations reported in the literature with rotation rates of hundreds and even $\sim2000 \dg$/day that occur on an intraday scale \cite{Ahnen2018}. To date only a few such events have been accidentally registered, since this requires dedicated high cadence monitoring programs \citep{Kiehlmann2021}. This kind of fast rotations are hard to explain by the model discussed in this paper, but they can be produced, e.g. by magnetic reconnection events \citep{Zhang2018,Hosking2020}.

Interestingly, the three EVPA rotation events visible in Fig.\ref{fig:all} appear to be very different. There are both clockwise and anti-clockwise rotations. Their rates and amplitudes differ significantly. Moreover, in the first event there is an acceleration of the rotation rate, while the other two events have stable rates. These differences can be explained by a presence of a secondary component(s) in the optical polarized emission of the jet. It is known that such components are common for blazars \citep{Holmes1984,Uemura2010} and they can dramatically change the appearance of the EVPA curve during rotations \citep{Cohen2020}. It is supported by our finding of the hidden EVPA rotation, which can also be explained by a strongly polarized constant component underlying the variable polarized emission. Another indication of a superposition of multiple polarized components in the optical emission is the dramatic variability of fractional polarization during considered EVPA rotation events from zero to dozens of percent. A detailed modeling of the EVPA curve is required in order to understand the evolution of polarization of the main emission feature itself.

One can notice in Fig.~\ref{fig:all} that there are many more radio knots during the considered time interval than detected pattern repetitions and EVPA rotations. If the system of emission regions in the jet sheath is persistent along the $\sim6$ years interval between the first and the last event, then other knots should have produced the gamma-ray pattern and EVPA rotations as well. We explain the lack of EVPA rotations associated with each knot by the two effects described above: by insufficient cadence of polarimetric observations and by the presence of multiple polarized emitters in the jet. The lack of the gamma-ray pattern repetition preceding each radio knot ejection can also be explained by a superposition of multiple gamma-ray emitting zones in the jet that camouflage the pattern with stronger flares. Additionally, the dissipation zone location and the EC scattering efficiency strongly depend on the moving emission feature bulk Lorentz factor \citep{Katarzynski2007}. Given the non-uniform density of the jet sheath it is possible that some of the emission features observed as radio knots dissipate most of their energy either at the beginning or at the end of the zone responsible for the gamma-ray flares pattern. Thereby, these knots can demonstrate only some of the flares a - c in the pattern, which would make it impossible to identify such events in the light curve.

Finally, we note that latest developments of observational techniques offer further observational tests of the model discussed in this work. It has been recently demonstrated that optical photocenters of AGN measured by {\em Gaia} space observatory have a systematic statistically significant tendency to be shifted either downstream or upstream the jet from the radio core \citep{Kovalev2017,Xu2021}. These VLBI - Gaia (VG) positional offsets are explained by presence of substructures that are astrometrically resolved by {\em Gaia} \citep{Plavin2019}. Moreover, it has been demonstrated that VG offsets are correlated with optical polarization properties of AGN \citep{Kovalev2020}. We speculate that events similar to those considered in this work can potentially produce a so called jitter of the optical photocenter in {\em Gaia} monitoring data. For instance, if knot C36 traveling with average $<\mu>=0.81$ mas yr$^{-1}$ \citep{Larionov2020} is indeed related to repeated pattern 3 that lasted for 236 days then the emission feature responsible for this event was located $\sim0.5$ mas upstream the core in the beginning of this pattern. Even if this emission feature constitutes a fraction of the total flux, it would be able to shift the optical photocenter by $\sim0.1-0.2$ mas \citep[see e.g.][]{Hwang2020}, which exceeds accuracy of a single {\em Gaia} measurement for 3C~279 in a bright state \citep{Mignard2010}. Existence of such jitter can be verified in future, when {\em Gaia} time series data become public. Currently, possible manifestation of such jitter can be seen in {\em Gaia} EDR3 \citep{Brown2021}, where 3C~279 has significant proper motion of $0.25\pm0.06$ mas/year with position angle $40\dg\pm14\dg$, which is parallel to the upstream jet direction within uncertainties. Additionally, 3C~279 has a significant $astrometric\_excess\_noise$ value of 0.2 mas reported, which exceeds the standard deviation of this quantity among object with comparable brightness ($\Delta Gmag<0.5$) within the 10 arcmin field around the blazar, and is the second largest in this field. Large and significant values of these two quantities can indicate a deviation of the source photocenter from the standard astrometric model, which is consistent with the discussed scenario. Furthermore, at least for brightest blazars in outbursts the expected a few tenths of milli-arcsecond separation between components with different polarization states in the optical band can be detected using polaroastrometry technique with large telescopes \citep{Safonov2015,Safonov2019}.

\section{Conclusions} \label{sec:concl}

We report on the discovery of a repeated pattern of gamma-ray flaring events in a blazar during EVPA rotation events. The sequence of observed events in the optical and gamma-ray bands allows us to conclude that at least in some blazars the long-term (weeks to months) EVPA rotations are unambiguously linked with propagation of emission features in the jet and are accompanied by a series of physically related gamma-ray flares.

\section{Data availability}

The gamma-ray data underlying this article are available in Harvard Dataverse, at \url{https://doi.org/10.7910/DVN/JDGTBD}.

\section*{Acknowledgements}
We thank E. Ros, C. Casadio, S. Kiehlmann and the anonymous reviewer for constructive comments that improved this manuscript. This work is based on Fermi data, obtained from Fermi Science Support Center, provided by NASA's Goddard Space Flight Center (GSFC). We acknowledge the hard work by the Fermi-LAT Collaboration that provided the community with unprecedented quality data and made Fermi Tools so readily available. This study was based in part on observations conducted using the 1.8 m Perkins Telescope Observatory (PTO) in Arizona, which is owned and operated by Boston University. Data from the Steward Observatory spectropolarimetric monitoring project were used. This program was supported by Fermi Guest Investigator grants NNX08AW56G, NNX09AU10G, NNX12AO93G, and NNX15AU81G. D.B. acknowledges support from the European Research Council (ERC) under the European Union Horizon 2020 research and innovation program under the grant agreement No 771282. The research at BU is partly supported by Fermi GI grants 80NSSC20K1567 and 80NSSC20K1566.

%%%%%%%%%%%%%%%%%%%%%%%%%%%%%%%%%%%%%%%%%%%%%%%%%%



%\bibliographystyle{mnras}

\bibliographystyle{mnras}
% Use the LaTeX power, use bibtex properly.
\bibliography{references}

%%%%%%%%%%%%%%%%%%%%%%%%%%%%%%%%%%%%%%%%%%%%%%%%%%
\appendix

\section{The gamma-ray curve parts in the original time scale} \label{app:A}

\begin{figure*}
   \centering
   \includegraphics[width=0.90\textwidth]{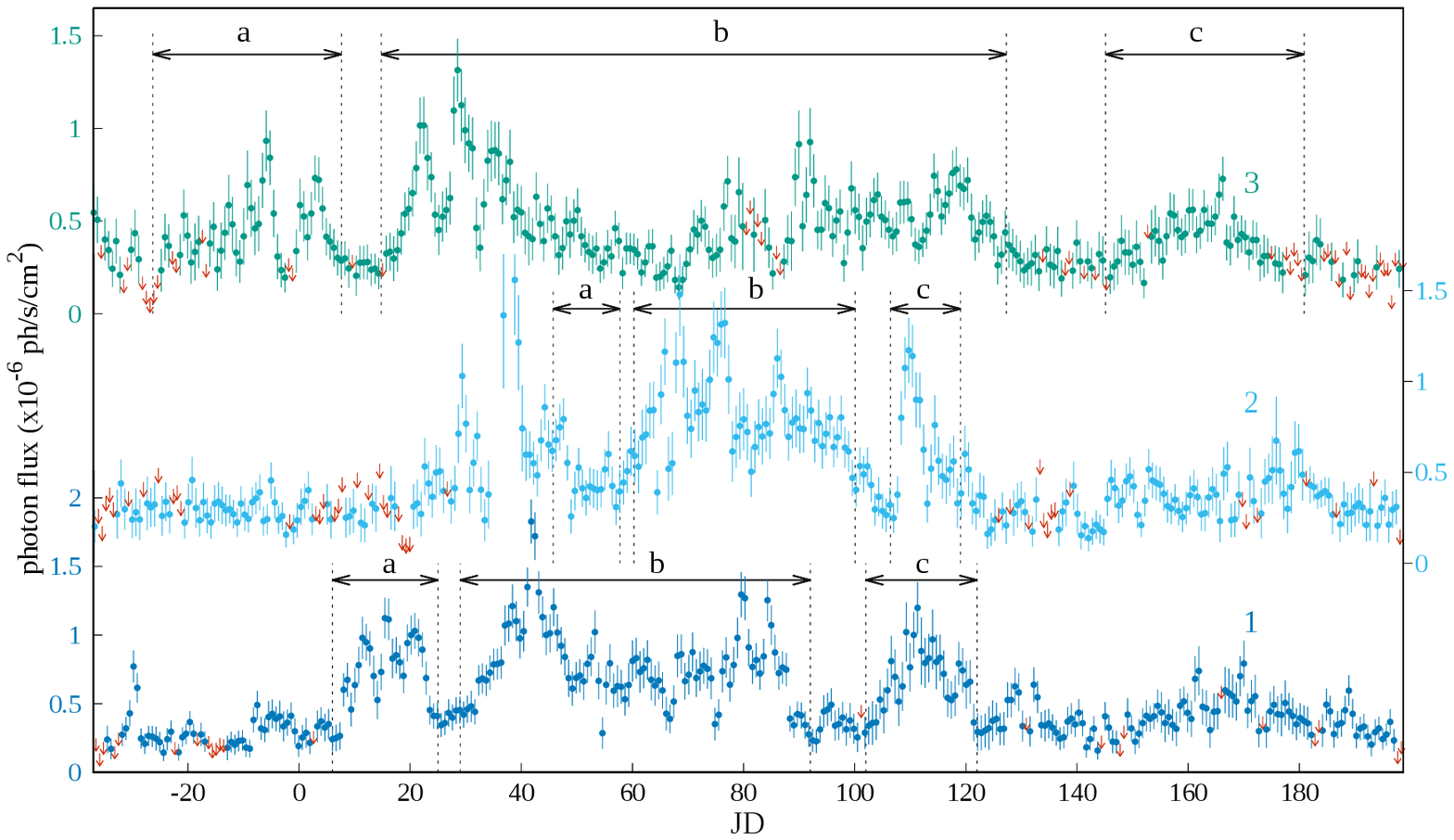}
      \caption{Gamma-ray light curves within periods denoted with dashed lines in Fig.~\ref{fig:all} in the original unaltered time scale. JD = 0 corresponds to MJD=55002 for the first curve. Curves 2 and 3 are shifted by -628 and -2315 days with respect to the first curve. Color coding and annotations are the same as in Fig.~\ref{fig:g_patt}.}
         \label{fig:origgamma}
\end{figure*}

Figure~\ref{fig:origgamma} demonstrates regions 1 -- 3 of the gamma-ray light curve in Fig.~\ref{fig:all} discussed along the paper in its original time scale, i.e. without squeezing and stretching of intervals 2 and 3.

\section{Similarity of the gamma-ray light curve regions} \label{app:B}

\begin{figure*}%f1
\includegraphics[width=\hsize]{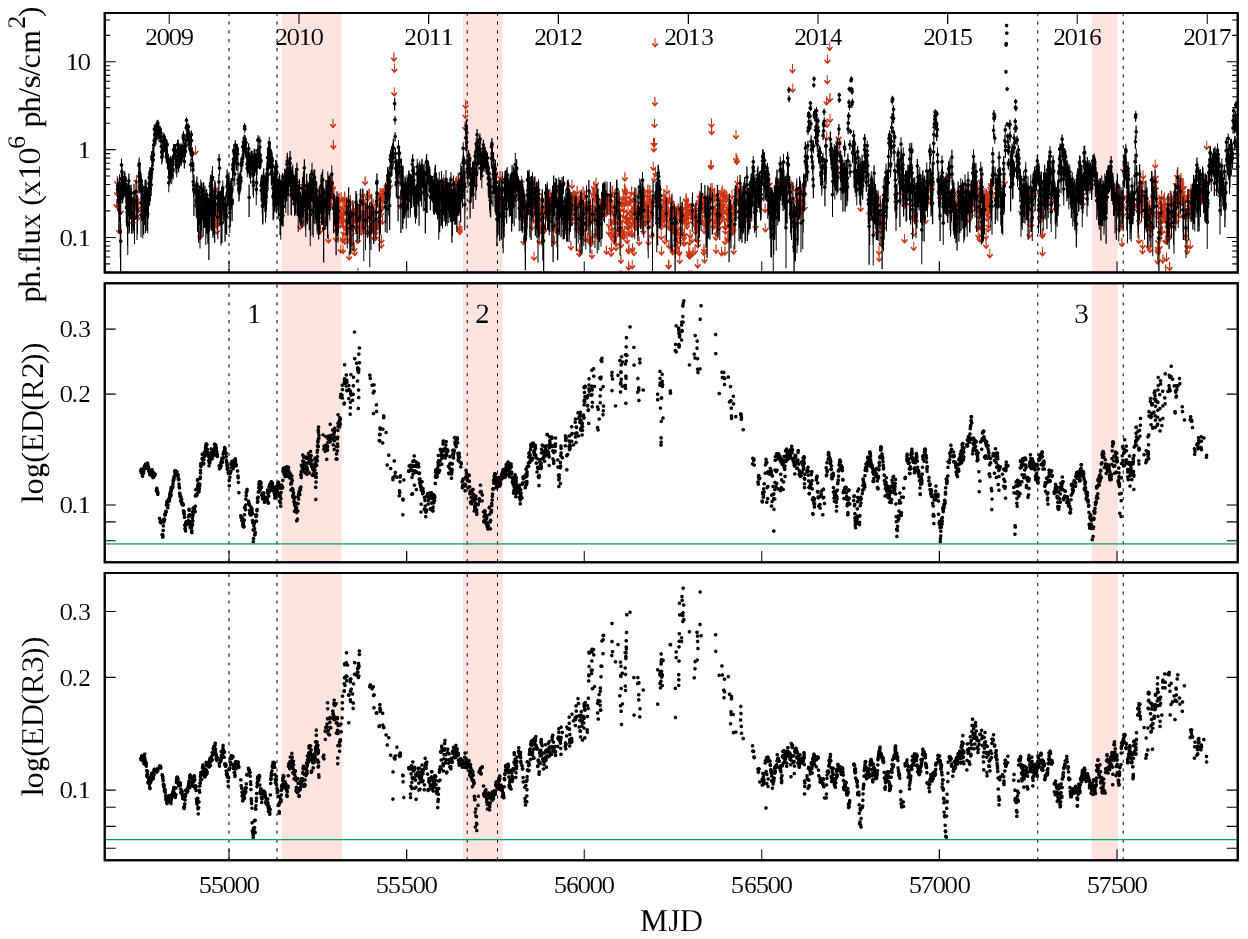}
\caption{Top panel: Gamma-ray light curve of 3C 279 with 1.35 days binning. Middle panel: Euclidean distance between the region 2 of the curve stretched by a factor of $\rtwofact$ and the entire gamma-ray light curve. Bottom panel: Euclidean distance between the region 3 of the curve compressed by a factor of $\rthreefact$ and the entire gamma-ray light curve. The pink areas mark periods of three largest EVPA rotations. The numbered intervals between the vertical dashed lines are the regions when the gamma-ray curve shows the repeated pattern. The green horizontal lines correspond to minimal values in the Euclidean distance curves, that are located in the middle of the region 1 in both cases.}
\label{fig:curve_match}
\end{figure*}

In this section we present mathematical evidence that the visual similarity of regions 1 -- 3 of the gamma-ray light curve of 3C 279 in Fig.~\ref{fig:g_patt} is not an illusion. There is a handful of measures proposed in order to assess similarity (distance) between light curves \citep[see e.g.][]{Lin2012}. Since the gamma-ray photon flux curves analyzed here are well-sampled and provide regular intervals between points, we use the most common and simple distance measure, namely the Euclidean distance. Assuming that light curves A and B are normalized and have the same length n, the Euclidean distance between them is defined as $D(A,B)=\sqrt{\sum_{i=1}^n (a_i-b_i)^2}$, where $a_i$ and $b_i$ are individual points of corresponding curves. In order to prove that region 1 provides the minimal Euclidean distance (i.e., the most similar) all along the curve to regions 2 and 3 after their time scale is transformed as described in sec.~\ref{sec:res} we performed the following exercise. In the curve with 1.35 days binning we selected only points corresponding to region 2 (55671.9 $\le$ MJD $\le$ 55755.4) and region 3 (57280.0 $\le$ MJD $\le$ 57515.7), which corresponds to the entire range of 0 $\le$ JD $\le$ 132 in Fig.~\ref{fig:g_patt}. Then MJD of selected points in regions 2 and 3 were multiplied by factors $\rtwofact$ and $\rthreefact$ respectively, following sec.~\ref{sec:res}. After this we iterated along the initial gamma-ray light curve calculating the Euclidean distance between it and the two light curve regions with transformed time scale. The result of this procedure is shown in the middle and bottom panels of Fig.~\ref{fig:curve_match}. For both regions 2 and 3 the minimal Euclidean distance is achieved in the middle of region 1. We repeated the same procedure using two other binnings of the light curve 1.4 and 2 d. Moreover, we used another distance measure TWED \citep{Lin2012} and varied MJD values limiting regions 2 and 3 within a few days. In all cases the result remains qualitatively persistent: region 1 gives the minimal or second minimal distance to regions 2 and 3 with transformed time scale. Therefore, we conclude that the similarity between considered intervals of the light curve is the best among all possible in the analyzed time range.

There is the opposite possibility that many random regions of the gamma-ray light curve under some transformation of the time scale can provide a relatively good similarity with region 1. Therefore, we estimated the probability that regions 2 and 3 under the found transformations are just accidentally close to region 1 in Euclidean distance. For this purpose we performed a Monte-Carlo simulation that selected a random piece of the entire light curve and a random time scale transformation factor. After transforming the selected region timescale using this factor and shifting it to the Julian Date range of the region 1, we calculated the Euclidean distance in the same way as before. In order to simulate regions 2 and 3 we ran this simulation separately allowing the timescale factor to vary 
in the ranges [1,~2] and [0.5,~1] respectively, so that the found factors $\rtwofact$ and $\rthreefact$ fall in these ranges. After repeating $10^6$ simulations for each of the two cases we found that a random region gives a better fit (smaller Euclidean distance) to region 1 in $<0.18$\% and $< 0.17$\% of trials for regions 2 and 3 respectively.

The two tests described in this section leave it beyond the reasonable doubt that the similarity of the gamma-ray light curve parts in regions 1, 2 and 3 is not random.

% Don't change these lines
\bsp	% typesetting comment
\label{lastpage}
\end{document}